\documentclass{article}
\usepackage{ijcai22}

\usepackage{times}
\usepackage{soul}
\usepackage{url}
\usepackage[hidelinks]{hyperref}
\usepackage[utf8]{inputenc}
\usepackage[small]{caption}
\usepackage{graphicx}
\usepackage{amsmath}
\usepackage{amsthm}
\usepackage{booktabs}
\usepackage{algorithm}
\usepackage{algorithmic}
\usepackage[T1]{fontenc}    
\usepackage{nicefrac}       
\usepackage{microtype}      

\usepackage{bm}
\usepackage{paralist}
\usepackage{array}
\usepackage{subcaption}
\usepackage[square]{natbib}
\usepackage{enumitem}

\newtheorem{example}{Example}
\newtheorem{theorem}{Theorem}

\newtheorem{definition}{Definition}
\newtheorem{lemma}{Lemma}

\newenvironment{proofof}[1]{{\vspace*{5pt} \noindent\bf Proof of #1:  }}{\hfill\rule{2mm}{2mm}\vspace*{5pt}}

\newcommand{\opt}{\textsf{OPT}}

\newcommand{\mbf}{\mathbf}

\newcommand{\algName}{\textsf{Patching}}
\newcommand{\mbfx}{\mathbf{x}}
\newcommand{\mbfr}{\mathbf{r}}
\newcommand{\mbfp}{\mathbf{p}}
\newcommand{\mstg}{(D, \mbfp)}

\newcommand{\SMaxTop}{\text{SharedMaxTop}}

\newcommand{\mbff}{\mathbf{f}}
\newcommand{\mbft}{\mathbf{t}}
\newcommand{\vmax}{V_\text{max}}
\newcommand{\tilmbfr}{\tilde{\mathbf{r}}}
\newcommand{\tilOmep}{\widetilde{\Omega}_p}
\newcommand{\findT}{\textsf{FindT}}
\newcommand{\maxtop}{\text{MaxTop}}
\newcommand{\cyclemt}{\text{CycleMaxTop}}

\title{Mixed Strategies for Security Games with General Defending Requirements\footnote{Xiaowei Wu is partially funded by FDCT (File no. 0143/2020/A3, SKL-IOTSC-2021-2023), the SRG of University of Macau (File no. SRG2020-00020-IOTSC) and GDST (2020B1212030003). Weijia Jia's work is supported by Guangdong Key Lab of AI and Multi-modal Data Processing, BNU-HKBU United International College (UIC), Zhuhai, No. 2020KSYS007; Chinese National Research Fund (NSFC), No. 61872239; Zhuhai Science-Tech Innovation Bureau, Nos. ZH22017001210119PWC and 28712217900001 and Guangdong Engineering Center for Artificial Intelligence and Future Education, Beijing Normal University, Zhuhai, Guangdong, China.}}

\begin{document}

\author{
Rufan Bai$^1$\and
Haoxing Lin$^2$\footnote{Part of this work was done while the author was working as an research assistant at the University of Macau.}\and
Xinyu Yang$^1$\and
Xiaowei Wu$^1$\and
Minming Li$^3$\And
Weijia Jia$^4$\\

\affiliations
$^1$IoTSC, University of Macau, $^2$National University of Singapore, $^3$City University of Hong Kong, $^4$BNU-UIC \& Beijing Normal University (Zhuhai)\\
\emails
yb97439@um.edu.mo,
haoxing.lin@comp.nus.edu.sg,
mb95466@um.edu.mo,\\
xiaoweiwu@um.edu.mo,
minming.li@cityu.edu.hk,
jiawj@uic.edu.cn
}

\maketitle

    \begin{abstract}
		The Stackelberg security game is played between a defender and an attacker, where the defender needs to allocate a limited amount of resources to multiple targets in order to minimize the loss due to adversarial attack by the attacker.
		While allowing targets to have different values, classic settings often assume uniform requirements to defend the targets.
		This enables existing results that study mixed strategies (randomized allocation algorithms) to adopt a \emph{compact representation} of the mixed strategies.
        
        In this work, we initiate the study of mixed strategies for the security games in which the targets can have different defending requirements.
        In contrast to the case of uniform defending requirement, for which an optimal mixed strategy can be computed efficiently, we show that computing the optimal mixed strategy is \textsf{NP}-hard for the general defending requirements setting.
        However, we show that strong upper and lower bounds for the optimal mixed strategy defending result can be derived.
        We propose an efficient close-to-optimal {\algName} algorithm that computes mixed strategies that use only few pure strategies.
        We also study the setting when the game is played on a network and resource sharing is enabled between neighboring targets.
		Our experimental results demonstrate the effectiveness of our algorithm in several large real-world datasets.
	\end{abstract}




\section{Introduction}
    
    Recently, security games have attracted much attention from the game theory society due to its applications in many real world scenarios~\citep{aamas/ShiehAYTBDMM12,aaai/Conitzer12,ijcai/FangST15}.
	Classical security games often model the problem as a Stackelberg game~\citep{aaai/NguyenYAKT13,ec/GanXGTRW19} that is played between two players, the \emph{defender} and the \emph{attacker}, where the defender is the \emph{leader} who commits to a defending strategy before the \emph{follower} (the attacker) observes and responds.
	In this paper we focus on zero-sum games~\citep{alshamsi2018optimal,aaai/TsaiNT12}, in which there are multiple targets to be defended, where each target $u$ has a value $\alpha_u$ that represents the loss if an attack at the target is successful, and a threshold $\theta_u$ that represents the resource needed to defend the attack, i.e., the defending requirement. If a target $u$ receives resource at least $\theta_u$, then no loss will occur if $u$ is under attack.
	In these games, the defender needs to decide an allocation of the limited resources to the targets; and the attacker will choose a target to attack after observing the strategy of the defender.
	The objective of the defender is to minimize the loss caused by the attack, which we refer to as the defending result of the allocation strategy.
	
	The allocation strategies are often categorized as \emph{pure strategies} and \emph{mixed strategies}. When the allocation is deterministic (resp. randomized), it is called a pure (resp. mixed) strategy.
	Formally speaking, a mixed strategy is a probability distribution over a set of pure strategies.
    It has been commonly observed that mixed strategies often achieve defending results that are much better than that of the best pure strategy.
    
    \begin{example}
        Consider the instance with targets $\{a,b,c,d\}$, each of which has threshold (defending requirement) equals $1$. The values of targets $\{a,b,c\}$ are $3$ and the value of target $d$ is 1.
        Given total resource $R = 2$, every pure strategy has defending result $3$ because there always exists a target among $\{ a,b,c \}$ with insufficient defending resource. 
        In contrast, the mixed strategy that applies each of the following three pure strategies $(1,1,0,0)$, $(1,0,1,0)$, and $(0,1,1,0)$ with probability $1/3$ achieves a defending result of $1$.
    \end{example}

	Most existing works that consider mixed strategies for security games assume that the thresholds of targets are uniform~\citep{ijcai/SinhaFAKT18,icc/NguyenAB09,atal/KiekintveldJTPOT09,aaai/KorzhykCP10}.
	This allows us to represent each mixed strategy by its corresponding \emph{compact representation}, in which the resources allocated to the targets are no longer binary.
	Instead, when a target receives resources below its threshold, it is assumed that the target is \emph{fractionally} defended when we evaluate the loss due to the attack.
	It has been shown by \cite{aaai/KorzhykCP10} that it takes polynomial time to translate any compact representation into a mixed strategy that uses $O(n^2)$ pure strategies and achieves the same result as the compact representation, where $n$ is the number of targets.
	
	In this paper, we consider the case when the thresholds of targets are general, i.e., the defending requirements of the targets can be different.
	For example, while defending against virus, consider the targets as cities and the resources as vaccines. The defending requirements to protect the cities naturally depend on the populations of cities, which can be dramatically different.
	Therefore, a natural question to ask is whether the {compact representation} still holds when targets have general thresholds. 
	In particular,
	
	\smallskip
	
	\emph{Can every compact representation be transformed into a mixed strategy that achieves the same defending result?}
	
	\smallskip
	
	Unfortunately, via the following example, we show that this is not true when targets have general thresholds.
	
	
	\begin{example}
	    Consider the instance with three targets with thresholds and values listed in the table below.
		Given total resource $R = 4$, it can be verified that the optimal mixed strategy applies each of the pure strategies $(3,0,1), (0,3,1)$ with probability $1/2$, and has defending result $1$. 
		However, the compact representation $(15/8, 15/8, 1/4)$ has defending result $3/4$ and clearly, there does not exist any mixed strategy that achieves the same defending result.
		\begin{table}[ht]
		    \footnotesize
			\linespread{1.2} 
			\centering
			\begin{tabular}{ c | c | c | c }
				\toprule
				Target & a & b & c \\
				\midrule
				Value & 2 & 2 & 1 \\
				\midrule
				Threshold & 3 & 3 & 1 \\
				\bottomrule
			\end{tabular}
			\vspace*{-5pt}
		\end{table}
	\end{example}

	

	\subsection{Our Contributions}
	
	Since compact representations and mixed strategies are no longer equivalent in the general threshold setting, in the following we refer to each compact representation as a \emph{fractional strategy}.
	Our work formalizes the mixed strategies and fractional strategies for security games with general defending requirements, and studies their connections and differences.
	
	\paragraph{Mixed vs. Fractional.}
	Our first contribution is a theoretical study to establish the connection between mixed and fractional strategies.
	We let $\opt_m(R)$ and $\opt_f(R)$ be the defending result of the optimal mixed and fractional strategy using total resource $R$, respectively. 
	We first show that computing the optimal mixed strategy is \textsf{NP}-hard, but we always have $\opt_m(R) \geq \opt_f(R)$. 
	Since the optimal fractional strategy with defending result $\opt_f(R)$ can be computed by solving a linear program, we use $\opt_f(R)$ to lower bound $\opt_m(R)$.
	More importantly, we show that given total resource $R$, we can always find a mixed strategy whose defending result is at most $\opt_f(R - \theta_\text{max})$, where $\theta_\text{max}$ is the maximum threshold of the nodes. 
	Moreover, we present a polynomial-time algorithm to compute the mixed strategy whose defending result is at most $\opt_f(R - \theta_\text{max})$, and guarantee that $O(n^2)$ pure strategies are used.
    By proving that the function $\opt_f(\cdot)$ is convex, we show that when $R$ is much larger than $\theta_\text{max}$, $\opt_m(R)$ and $\opt_f(R)$ are very close to each other.
    To the best of our knowledge, we are the first to theoretically establish the almost-equivalence between mixed and fractional strategies for the security game with general defending requirements.

    \paragraph{Algorithm with Small Support.}
	For practical use purpose, mixed strategies that use few pure strategies are often preferred, e.g., it is unrealistic to deploy a mixed strategy that uses $\omega(n)$ pure strategies when the number of targets $n$ is large.
	Thus we study the computation of mixed strategies that use few pure strategies, i.e., those with small \emph{supports}.
	Motivated by the Double Oracle algorithm by Jain et al.~\citep{aamas/JainKVCPT11} and the column generation techniques~\citep{atal/JainCT13,aaai/WangYA16,aaai/GanAVG17}, we propose the \algName\space algorithm that in each iteration finds and includes a new pure strategy to patch the nodes that are poorly defended by the current mixed strategy.
	We show that given a bounded size pure strategy set $D$, our algorithm takes time polynomial in $|D|$ to compute an optimal mixed strategy with support $D$.

	\paragraph{Resource Sharing.}
	We also study the setting with resource sharing in the network, which is motivated by patrolling and surveillance camera installation applications~\citep{aips/VorobeychikATS14, ijcai/YinXGAJ15, aaai/BaiLYW21}. 
	In this setting the targets are represented by the nodes of a network, and when a certain target is attacked, some fraction of resources allocated to its neighbors can be shared to the target.
	Similar to ours, \cite{aaai/LiTW20} consider the model with general thresholds, but they only study the computation of pure strategies.
	When resource sharing is allowed, we show that the gap between $\opt_m(R)$ and $\opt_f(R)$ can be arbitrarily large.
	Therefore, the idea of rounding fractional strategies to get mixed strategies with good approximation guarantee is no longer feasible.
	However, our {\algName} algorithm can still be applied to compute mixed strategies efficiently in the resource sharing setting.
	We show that under certain conditions, the algorithm is able to make progress towards decreasing the defending result.
	
	\paragraph{Experiments.}
	Finally, we conduct extensive experiments on several large real-world datasets to verify our analysis and test our algorithms.
	The experimental results show that the \algName\space algorithm efficiently computes mixed strategies with small support, e.g., using $5$ pure strategies, whose defending result dramatically improves the optimal pure strategies, and is close-to-optimal in many cases.
	
	\smallskip
	
	The rest of the paper is structured as follows. Section~\ref{sec:prelim} introduces a formal description of the models. Section~\ref{ssec:computation-pure-frac} to \ref{ssec:upperbound} focus on the relation between mixed and fractional strategy, where we establish most of our theoretical results. Section~\ref{ssec:hardness-sharing} shows some negative results in the resource sharing setting. The {\algName} algorithm is presented in Section~\ref{sec:algorithm} and the experimental results are included in Section~\ref{sec:experiments}.

	\subsection{Other Related Works}
	
	There are works that consider the security game with resource sharing as a dynamic process in which it takes time for the neighboring nodes to share the resources~\citep{atal/BasilicoGA09,ijcai/YinXGAJ15}.
	Motivated by the applications to stop virus from spreading, network security games with contagious attacks have also received a considerable attention in recent years~\citep{jcss/AspnesCY06,icdcs/KumarRSS10,aaai/TsaiNT12,jet/AcemogluMO16,aaai/BaiLYW21}.
	In these models the attack at a node can spread to its neighbors, and the loss is evaluated at all nodes under attack. 
	There are also works that consider multi-defender games, where each defender is responsible for one target~\citep{ijcai/LouV15,atal/GanEW18,atal/GanEKW20}.

    \section{Preliminaries} \label{sec:prelim}
	
	In this section we present the model we study. We define our model in the most general form, i.e., including the network structure and with resource sharing, and consider the model without resource sharing as a restricted setting.
	
	We model the network as an undirected connected graph $G(V,E)$, where each node $u \in V$ has a \emph{threshold} $\theta_u$ that represents the defending requirement, and a \emph{value} $\alpha_u$ that represents the possible damage due to an attack at node $u$.
	Each edge $e\in E$ is associated with a weight $w_{uv}$, which represents the efficiency of resource sharing between the two endpoints.
	We use $N(u):= \{v\in V: (u,v)\in E\}$ to denote the set of neighbors for node $u\in V$.
	We use $n$ and $m$ to denote the number of nodes and edges in the graph $G$, respectively.
	For any integer $i$, we use $[i]$ to denote $\{1,2,\ldots,i\}$.
	
	
	The defender has a total resource of $R$ that can be distributed to nodes in $V$.
	We use $r_u$ to denote the \emph{defending resource}\footnote{As in~\citep{aaai/LiTW20,aaai/BaiLYW21}, we assume the resource can be allocated arbitrarily in our model.} allocated to node $u$.
	Thus we require $\sum_{u\in V}r_u \leq R$.
	
	\begin{definition}[Pure Strategy]
		We use $\mathbf{r} = \{ r_u \}_{u\in V}$ to denote a \emph{pure strategy} and $\Omega_{p}(R) = \{ \mathbf{r}\in [0, R]^V: \| \mbf{r} \| = \sum_{u\in V} r_u  \leq R \}$ \footnote{Throughout this paper we use $\|\cdot\|$ to denote the $L_1$ norm.}
		to denote the collection of pure strategies using resource $R$.
	    When $R$ is clear from the context, we use $\Omega_p$ to denote $\Omega_p(R)$.
	\end{definition}
	
	We consider resource sharing in our model.
	That is, when node $u$ is under attack, it can receive $w_{uv}\cdot r_v$ units of resource shared from each of its neighbors $v\in N(u)$.
	
	\begin{definition}[Defending Power]
		Given pure strategy $\mbf{r}$, the defending power of node $u$ is defined as $\pi_u(\mbf{r}) = r_u + \sum_{v\in N(u)} w_{uv} \cdot r_v$.
		We use $\bm{\pi}(\mbf{r}) = (\pi_u(\mbf{r}))_{u\in V}$ to denote defending powers of nodes.
	\end{definition}
	
	\begin{definition}[Defending Status]
		Given a pure strategy $\mathbf{r}$, we use $\mathbf{x}(\mathbf{r})\in\{0, 1\}^V$ to denote the defending status of the nodes under $\mathbf{r}$, where each node $u$ has $x_u(\mathbf{r}) = 1$ if $\pi_u(\mbf{r}) \geq \theta_u$, i.e., node $u$ is well defended; and $x_u(\mathbf{r}) = 0$ otherwise.
	\end{definition}
	
	Each pure strategy $\mbf{r}$ has a unique defending status $\mbf{x}(\mbf{r})$ but different strategies can have the same defending status.
	
	\begin{definition}[Defending Result]
		Given a pure strategy $\mathbf{r}$, when node $u\in V$ is under attack, the loss is given by $L_p(u, \mbf{r}) = \alpha_u$ if $x_u(\mbf{r}) = 0$; $L_p(u, \mbf{r}) = 0$ otherwise.
		The defending result of strategy $\mbf{r}$ is defined as the maximum loss due to an attack: $L_{p}(\mathbf{r}) = \max_{u\in V} \{ L_p(u,\mathbf{r}) \}$.
	\end{definition}
	
	We use $\mbf{r}^*$ to denote the optimal pure strategy, i.e., the pure strategy that has the minimum defending result $\mathbf{r}^* = \arg\min_{\mathbf{r}\in \Omega_{p}} \{ L_p(\mathbf{r}) \}$.
	The corresponding defending result is defined as $\opt_p = L_p(\mathbf{r}^*)$.
	
	\begin{definition}[Mixed Strategy]
		A mixed strategy is denoted by $(D, \mbf{p})$, where $D\subseteq \Omega_p$ is a subset of pure strategies and $\mbf{p}$ is a probability distribution over $D$.
		For each $\mbf{r}\in D$, we use $p(\mbf{r})$ to denote the probability that pure strategy $\mbf{r}$ is used.
	\end{definition}
	
	A mixed strategy is a randomized algorithm that applies pure strategies with certain probabilities.
	Note that $\sum_{\mbf{r}\in D} p(\mbf{r}) = 1$.
	We can also interpret $\mbf{p}$ as a $|D|$ dimension vector with $\| \mbf{p} \| = 1$.
	We use ${\Omega}_{m}(R) = \{(D, \mathbf{p}): D \subseteq \Omega_{p}(R), \mathbf{p}\in [0,1]^{|D|}, \|\mathbf{p}\| = 1 \}$ to denote the collection of all mixed strategies using resource $R$.
%
	When $R$ is clear from the context, we use $\Omega_m$ to denote $\Omega_m(R)$.
	
	\begin{definition}[Defending Status of Mixed Strategy]
		Given a mixed strategy $(D, \mbf{p})$, we use $x_u(D,\mbf{p}) = \sum_{\mbf{r}\in D} p(\mbf{r})\cdot x_u(\mbf{r})$ to denote the defending status of node $u\in V$ under the mixed strategy $(D, \mbf{p})$.
		In other words, $x_u(D,\mbf{p})$ is the probability that node $u$ is well defended under mixed strategy $(D,\mbf{p})$.
		We use $\mathbf{x}(D,\mbf{p}) = (x_u(D,\mbf{p}))_{u\in V}\in [0,1]^V$ to denote the defending status of $(D, \mbf{p})$.
	\end{definition}
	
	\begin{definition}[Defending Result of Mixed Strategy]
		Given mixed strategy $(D, \mathbf{p})$, we use $L_m(u, (D, \mbf{p})) = (1 - x_u(D,\mbf{p})) \cdot \alpha_u$ to denote the (expected) loss when node $u$ is under attack.
		The defending result is defined as $L_{m}(D, \mathbf{p}) = \max_{u\in V} \{ L_{m}(u,(D, \mathbf{p})) \}$.
	\end{definition}
	
	We use $(D^*, \mathbf{p}^*)$ to denote the optimal mixed strategy, i.e., the mixed strategy with the minimum defending result. The corresponding defending result is defined as $\opt_m = L_m(D^*,\mbf{p}^*)$.
	
	Next we define the fractional strategies.
	Technically, a fractional strategy is not a strategy, but instead a pure strategy equipped with a fractional valuation of defending loss.
	In the remaining of this paper, when a pure strategy is evaluated by its fractional loss, we call it a \emph{fractional strategy}.
	
	\begin{definition}[Fractional Loss]
		Given a pure strategy $\mbf{r} \in \Omega_p$, we evaluate the fractional loss when node $u$ is attacked by $L_f(u, \mbf{r}) = (1- \min\{\pi_u(\mbf{r})/\theta_u, 1\}) \cdot \alpha_u$.
		The fractional defending result is defined as $L_f(\mbf{r}) = \max_{u\in V} \{ L_f(u,\mbf{r}) \}$.
	\end{definition}
	
	In a fractional strategy, if a node $u$ has defending power $\pi_u$, then we assume that $\min\{\pi_u/\theta_u, 1\}$ fraction of the node is defended.
	Thus when node $u$ is under attack, the loss is given by $(1- \min\{\pi_u/\theta_u, 1\}) \cdot \alpha_u$.
	We use $\mathbf{\tilde{r}}^*$ to denote the optimal fractional strategy, i.e., the strategy with minimum $L_{f}(\mathbf{\tilde{r}^*})$.
	The corresponding defending result is defined as $\opt_f$.
	We use $\opt_p(R),\opt_m(R)$ and $\opt_f(R)$ to denote the defending result of the optimal pure, mixed and fractional strategy using total resource $R$, respectively.
	When $R$ is clear from the context, we simply use  $\opt_p,\opt_m$ and $\opt_f$.
	The following lemma implies that the optimal fractional strategy has a defending result at most that of the optimal mixed strategy.
	
	\begin{lemma}\label{lemma:relation}
		For any given problem instance, we have $\opt_p \geq \opt_m 
		\geq \opt_f$.
	\end{lemma}
	\begin{proof}
		Note that every pure strategy $\mbf{r}$ is also a mixed strategy (with $D = \{ \mbf{r} \}$ and $p(\mbf{r}) = 1$).
		Hence the first inequality trivially holds.
		In the following, we show that for any mixed strategy $(D, \mathbf{p})$, we can find a fractional strategy $\mathbf{\tilde{r}}$ using the same total resource $R$ such that $L_{m}(D, \mathbf{p}) \geq L_{f}(\mathbf{\tilde{r}})$.
		
		Let $\tilde{r}_u = \sum_{\mbf{r}\in D} p(\mbf{r})\cdot r_u$ be the expected resource node $u$ receives under $(D, \mathbf{p})$. Let $\mbf{\tilde{r}} = (\tilde{r}_u)_{u\in V}$ be the resulting fractional strategy.
		Note that we have $\pi_u(\mbf{\tilde{r}}) = \sum_{\mbf{r}\in D} p(\mbf{r})\cdot \pi_u(\mbf{r})$.
		Furthermore, $\| \mbf{\tilde{r}} \|\leq R$, i.e., it uses a total resource at most $R$.
		Observe that when node $u$ is under attack we have
		\begin{align*} 
			& L_{m}(u, (D, \mathbf{p})) = (1 - x_u(D,\mbf{p})) \cdot \alpha_u \\
			= & \textstyle (1 - \sum_{\mbf{r}\in D} p(\mbf{r})\cdot x_u(\mbf{r}) ) \cdot \alpha_u \\
			\geq & \textstyle (1 - \sum_{\mbf{r}\in D} p(\mbf{r})\cdot \min\{\pi_u(\mbf{r})/\theta_u, 1\} ) \cdot \alpha_u \\
			\geq & \textstyle (1 - \min\{ \sum_{\mbf{r}\in D} p(\mbf{r})\cdot \pi_u(\mbf{r})/\theta_u, 1\} ) \cdot \alpha_u
			= L_f(u, \mbf{\tilde{r}}).
		\end{align*}
		
		It means that $L_{m}(D, \mathbf{p})\geq L_{f}(\mathbf{\tilde{r}})$ since above relation holds for each node, which implies that $\opt_{m} \geq \opt_{f}$.
	\end{proof}

    \section{Computation of Strategies} \label{sec:computation}
	
	In this section we consider the computation of the optimal pure, mixed and fractional strategies, and analyze some properties regarding the optimal defending results of different strategies.
	
	\subsection{Optimal Pure and Fractional Strategy} \label{ssec:computation-pure-frac}
	
	We remark that our model is equal to the ``single threshold'' model of~\cite{aaai/LiTW20}.
	We thus use their algorithm (that runs in polynomial time) to compute an optimal pure strategy.
	Roughly speaking, in their algorithm a target defending result $\alpha$ is fixed and the goal is to decide whether it is possible to defend all nodes $A(\alpha) = \{ u\in V: \alpha_u > \alpha \} $ with value larger than $\alpha$.
	For every fixed $\alpha$ the above decision problem can be solved by computing a feasibility LP with constraints $\sum_{u\in V} r_u \leq R$ and $ r_u  +  \sum_{v\in N(u)} w_{uv} \cdot r_v \geq \theta_u$ for every $u\in A(\alpha)$.
	Combining the above sub-routine with a binary search on $\alpha \in \{ \alpha_u \}_{u\in V}\cup\{0\}$ yields a polynomial time algorithm for computing the optimal pure strategy, i.e., with the minimum achievable defending result $\alpha$.
	
	The computation of the optimal fractional strategy can be done efficiently by solving the following linear program $(\textsf{LP}_f(R))$, where we introduce a variable $r_u$ for each node $u\in V$ that represents the resource $u$ receives, and a variable $L$ for the defending result.
	\begin{align*}
		(\textsf{LP}_f(R)) \qquad
		\text{minimize} \;\quad\qquad \qquad L\quad & \\
		\text{subject to} \qquad\quad \textstyle \sum_{u\in V} r_u &\leq R,   \\
		\textstyle (1-(r_u  + \textstyle \sum_{v\in N(u)} w_{uv} \cdot r_v )/\theta_u ) \cdot \alpha_u &\leq L, \qquad \forall u \in V 
	\end{align*}
	
	By solving the above LP we can get the optimal fractional strategy $\mbf{\tilde{r}}^*$, whose defending result $\opt_f$ is the optimal objective of the LP. 
	From Lemma \ref{lemma:relation}, we have $\opt_{f} \leq \opt_{m}$, i.e., we can use $\opt_f$ as a lower bound for the defending result of the optimal mixed strategy (which is \textsf{NP}-hard to compute, as we will show in the next subsection).
	In the following we show that the optimal objective of the above LP is a convex function of the total resource $R$.

	\begin{lemma}[Convexity]\label{lemma:convexity}
		Given resource $R_1$ and $R_2$, we have
		\begin{equation*}
		\textstyle \opt_f(R_1) + \opt_f(R_2) \geq 2\cdot \opt_f\left(\frac{1}{2}(R_1+R_2)\right).
		\end{equation*} 
	\end{lemma}
	\begin{proof}
        Let $\mbf{\tilde{r}_1}^*$ and $\mbf{\tilde{r}_2}^*$ be the optimal fractional strategies given resource $R_1$ and $R_2$, respectively.
    	Note that $(\mbf{\tilde{r}_1}^*, \opt_f(R_1))$ and $(\mbf{\tilde{r}_2}^*, \opt_f(R_2))$ are feasible solutions to $(\textsf{LP}_f(R_1))$ and $(\textsf{LP}_f(R_2))$, respectively.
    	Let $\mbf{\tilde{r}} = \frac{1}{2}\cdot (\mbf{\tilde{r}_1}^* + \mbf{\tilde{r}_2}^*)$.
    	In the following we show that $(\mbf{\tilde{r}}, \frac{1}{2}\cdot (\opt_f(R_1) + \opt_f(R_2)) )$ is a feasible solution to $(\textsf{LP}_f(\frac{1}{2}(R_1+ R_2) ) )$.
    	The first constraint of the LP trivially holds because $\| \mbf{\tilde{r}} \| = \frac{1}{2}\cdot (\| \mbf{\tilde{r}}^*_1 \| + \| \mbf{\tilde{r}}^*_2 \|) \leq \frac{1}{2}(R_1 + R_2)$.
    	By the feasibility of $(\mbf{\tilde{r}_1}^*, \opt_f(R_1))$ and $(\mbf{\tilde{r}_2}^*, \opt_f(R_2))$, we have the following relations:
    	\begin{align*}
    	(1 - \pi_u(\mbf{\tilde{r}_1}^*)/\theta_u ) \cdot \alpha_u \leq \opt_f(R_1), \qquad \forall u \in V \\
    	(1 - \pi_u(\mbf{\tilde{r}_2}^*)/\theta_u ) \cdot \alpha_u \leq \opt_f(R_2), \qquad \forall u \in V.
    	\end{align*}
    	
    	Combining the two sets of inequalities we get
    	\begin{align*}
    	\textstyle &(1 - \frac{1}{2}( \pi_u(\mbf{\tilde{r}_1^*}) + \pi_u(\mbf{\tilde{r}_2^*}) )/\theta_u ) \cdot \alpha_u \\
    	\leq  &\frac{1}{2}\cdot (\opt_f(R_1) + \opt_f(R_2)),  \forall u \in V.
    	\end{align*}
    	
    	Since we have $\pi_u(\mbf{\tilde{r}}) = \frac{1}{2}( \pi_u(\mbf{\tilde{r}_1^*}) + \pi_u(\mbf{\tilde{r}_2^*}) )$, we conclude that $(\mbf{\tilde{r}}, \frac{1}{2} (\opt_f(R_1) + \opt_f(R_2)) )$ is a feasible solution to $(\textsf{LP}_f(\frac{1}{2}(R_1+ R_2) ) )$.
    	Consequently the optimal objective of the LP has
    	\begin{equation*}
    	\textstyle \opt_f(\frac{1}{2}(R_1 + R_2)) \leq  \frac{1}{2}\cdot (\opt_f(R_1) + \opt_f(R_2)).
    	\end{equation*}		
    	
    	Rearranging the inequality concludes the proof.
    \end{proof}

	\subsection{Hardness for Computing Mixed Strategies}
	
	We have shown that the optimal pure and fractional strategies can be computed efficiently.
	Unfortunately, we show that computing the optimal mixed strategy is \textsf{NP}-hard, even in the \emph{isolated model}, i.e., when $w_{uv} = 0$ for all $(u,v)\in E$.
	
	\begin{theorem}\label{thy:hardness}
		Unless $P = NP$, there does not exist any polynomial time algorithm that given a graph $G(V,E)$ and resource $R$ computes the optimal mixed strategy, even under the isolated model.
	\end{theorem}
	\begin{proof}
		We prove the hardness result by a reduction from the Even Partition problem, which is known to be \textsf{NP}-complete~\citep{gent1996phase}.
		Given a set of numbers $A = \{a_1, a_2, ..., a_n\}$, the problem is to decide whether $A$ can be partitioned into two subsets of equal sum.
		Given set $A$, we construct the instance of the defending problem as follows.
		Let $G(V,E)$ be a graph with $V = [n]$ and $E = \emptyset$.
		For each node $i\in V$ we set $\theta_i = a_i$ and $\alpha_i = 1$.
		We set the total resource $R = \frac{1}{2}\sum_{i\in A} a_i$.
		
		Obviously if $A$ can be partitioned into two sets $A_1$ and $A_2$ of equal sum, then both of them have sum equals to $R$.
		Then we can define two pure strategies: the first strategy allocates resource $r_i = \theta_i$ for each $i\in A_1$; the second one allocates resource $r_i = \theta_i$ for each $i\in A_2$.
		Then we define a mixed strategy that applies each of these two strategies with probability $0.5$. 
		It is easy to check that the defending result is $0.5$ since the defending status of each node is $0.5$.
		Hence if $A$ has an even partition, we have $\opt_m(R) \leq 0.5$.
		
		On the other hand, we show that if $A$ does not have an even partition, then $\opt_m(R) > 0.5$.
		Let $R'$ be the maximum sum of numbers in $A$ that is at most $R$.
		Since $A$ does not have an even partition, we have $R' < R$.
		Moreover, in every pure strategy the total threshold of nodes that are well defended is at most $R'$.
		In other words, for every $\mbfr\in \Omega_p(R)$, there is a corresponding $\mbfr' \in \Omega_p(R')$ with $\mbfx(\mbfr') = \mbfx(\mbfr)$.
		Thus we have $\opt_m(R) = \opt_m(R')$. 	
		Observe that since $R' < R=\frac{1}{2}\cdot\sum_{i\in V}\theta_i$, in any fractional strategy using total resource $R'$, there must exist a node $i\in V$ with $r_i < 0.5\cdot \theta_i$.
		Consequently, we have $\opt_f(R') > 0.5$.
		Finally, by Lemma~\ref{lemma:relation}, we have $\opt_m(R) = \opt_m(R') \geq \opt_f(R') > 0.5$, as claimed.
		
		In conclusion, we have $\opt_m(R) \leq 0.5$ if and only if $A$ admits an even partition. Since the reduction is in polynomial time, we know that the computation of the optimal mixed strategy is \textsf{NP}-hard.
	\end{proof}

	\subsection{A Strong Upper Bound for Isolated Model}\label{ssec:upperbound}

	While computing the optimal mixed strategy is \textsf{NP}-hard, we can use $\opt_f(R)$ to give a lower bound on $\opt_m(R)$.
	In other words, if a mixed strategy has a defending result close to $\opt_f(R)$, then it is close-to-optimal.
	However, if the lower bound is loose, then no such mixed strategy exists.
	Therefore, it is crucial to know whether this lower bound is tight.
	In this section, we show that in the isolated model, we can give a strong upper bound on $\opt_m(R)$, which shows that $\opt_f(R)$ is an almost tight lower bound when $R$ is large.
	
	\begin{theorem}\label{theorem:upper-bound-isolated}
		In the isolated model, given any instance $G(V,E)$ and a total resource $R$, we have
		\begin{equation*}
		\opt_m(R) \leq \opt_f(R-\theta_\text{max}),
		\end{equation*}
		where $\theta_\text{max} = \max_{u\in V} \{ \theta_u \}$ is the maximum threshold of the nodes.
	\end{theorem}
	
	Before presenting the proof, we remark that by convexity of the function $\opt_f(\cdot)$, we have
	\begin{align*}
	&\opt_f(R-\theta_\text{max}) \leq \textstyle \frac{R - \theta_\text{max}}{R}\cdot \opt_f(R) + \frac{\theta_\text{max}}{R}\cdot \opt_f(0) \\
	= & \textstyle \opt_f(R) \!+\! \frac{\theta_\text{max}}{R}\cdot ( \max_{u\in V}\{ \alpha_u \} \!-\! \opt_f(R) ).
	\end{align*}
	
	In other words, when $R \gg \theta_\text{max}$, $\opt_f(R)$ and $\opt_f(R-\theta_\text{max})$ have very similar values.
	Hence combining Lemma~\ref{lemma:relation} and Theorem~\ref{theorem:upper-bound-isolated}, we have strong upper and lower bounds on $\opt_m(R)$ when $R \gg \theta_\text{max}$.
	Furthermore, we remark that following our analysis, it can be verified that if $\theta_u = \theta_\text{max}$ for all nodes $u\in V$ and $R$ is divisible by $\theta_\text{max}$, then we can prove the stronger result $\opt_m(R) = \opt_f(R)$. Moreover, there exists a mixed strategy $(D,\mbfp)$ with $|D| = O(n^2)$ that achieves this defending result.
	In other words, our analysis also reproduces the result of~\citep{aaai/KorzhykCP10}.
	We prove Theorem~\ref{theorem:upper-bound-isolated} by showing the following lemma.
	
	\begin{lemma}\label{lemma:from-f-to-mixed}
		Given any vector $\mbff \in [0,1]^V$ with $\sum_{u\in V} f_u\cdot \theta_u \leq R-\theta_\text{max}$, we can compute in polynomial time a mixed strategy $(D,\mbfp)\in \Omega_{m}(R)$ with $|D|=O(n^2)$ such that $\mbfx(D,\mbfp) = \mbff$.
	\end{lemma}

In particular, let $\tilmbfr^*$ be the optimal fractional strategy using resource $R-\theta_\text{max}$.
Note that in the isolated model we have $\pi_u(\tilmbfr^*) = \tilde{r}^*_u$.
Let $\mbff \in [0,1]^V$ be defined by $f_u = \min\{  \tilde{r}^*_u / \theta_u, 1 \}$, for all $u\in V$.
That is, $f_u$ is the fraction node $u$ is defended in the fractional strategy.
Then $\mbff$ satisfies the condition of Lemma~\ref{lemma:from-f-to-mixed}, and hence there exists a mixed strategy $(D,\mbfp)\in \Omega_{m}(R)$ with $\mbfx(D,\mbfp) = \mbff$.
Hence we have
\begin{align*}
&\opt_m(R) \leq L_m(D,\mbfp) = \max_{u\in V} \{ ( 1 - x_u(D,\mbfp) )\cdot \alpha_u \} \\
= &\max_{u\in V} \{ ( 1 - f_u )\cdot \alpha_u \} = L_f(\tilmbfr^*) = \opt_f(R-\theta_\text{max}).
\end{align*}
	
\subsection{Proof of Lemma~\ref{lemma:from-f-to-mixed}}

We prove the lemma by giving a polynomial time algorithm that given the vector $\mbff$ computes the mixed strategy $(D,\mbfp)$ with the claimed properties.
For convenience of discussion, we first introduce the following notations.

\paragraph{Notations.}
In the isolated model, it makes no sense to allocate resource $r_u \in (0,\theta_u)$ to a node $u$.
Thus we only consider pure strategies $\mbfr$ with $r_u\in \{0, \theta_u\}$ for all $u\in V$, and let $\tilOmep(R)$ be the collection of such pure strategies using total resource at most $R$.
For a vector $\mbff\in [0,1]^V$, we use $\vmax(\mbff) = \{ u \in V: f_u = \max_{v \in V} \{f_v\} \}$ to denote the set of nodes $u$ with maximum $f_u$, and $V_0(\mbff) = \{u \in V: f_u = 0\}$.
In addition, given vector $\mbff$, we define $\maxtop(\mbff) \subseteq V$ to be a set of nodes with total threshold at most $R$ as follows.
We initialize $\maxtop(\mbff) \leftarrow \emptyset$ and then greedily include nodes $u\in V\setminus \maxtop(\mbff)$ with maximum $f_u$ value (break ties by the index of nodes) into $\maxtop(\mbff)$ as long as $f_u > 0$ and the resulting set of nodes has total threshold at most $R$.

\begin{algorithm}[ht]
	\caption{MaxTop($\mbff$)}
	\label{alg:maxtop}
	\textbf{Input}: the maximum \\
    \textbf{Output}: mixed strategy set $\mstg$ 
	\begin{algorithmic}[1]
	\STATE initialize $M \leftarrow \emptyset$ 
	\WHILE{ $M \neq V\setminus V_0(\mbff) $ }
		\STATE $u \leftarrow \arg\max_{v\in V\setminus M} \{ f_v \} $ 
		\IF{$\theta_u + \sum_{v\in M} \theta_v \leq R$}
			\STATE $M \leftarrow M\cup \{ u \}$ 
		\ELSE
			\STATE \textbf{return} $M$ 
		\ENDIF
	\ENDWHILE
    \end{algorithmic}
\end{algorithm}

Notice that there exists a pure strategy $\mbfr \in \tilOmep(R)$ that defends (and only defends) the nodes in $\maxtop(\mbff)$ simultaneously.
Furthermore, unless $\maxtop(\mbff)$ contains all nodes with non-zero $f$ values, i.e., $M = V\setminus V_0(\mbff)$, the total resource $\mbfr$ uses is $\| \mbfr \| = \sum_{u\in \maxtop(\mbff)} \theta_u > R - \theta_\text{max}$.

\subsubsection{Overview} \label{sssec:overview}

Let $\mbff^{(0)}$ be the vector $\mbff$ given in Lemma~\ref{lemma:from-f-to-mixed}.
Recall that our goal is to find $O(n^2)$ pure strategies $D\subseteq \tilOmep(R)$ and associate a probability $p(\mbfr)$ to each $\mbfr\in D$ satisfying \footnote{Formally, we need equality here. However, given any mixed strategy whose total probability of pure strategies is $1-\epsilon$, we can add a dummy pure strategy (that allocates $0$ resource to every node) with probability $\epsilon$ without changing the defending result.}
\begin{equation}
\textstyle \sum_{\mbfr\in D} p(\mbfr) \leq 1, \label{eq:prob-constraint}
\end{equation}
such that $\mbfx(D,\mbfp) = \mbff^{(0)}$.
We implement this goal by progressively including new pure strategies (with certain probabilities) into $D$ under the setting that
\begin{equation}
\mbfx(D,\mbfp) \leq \mbff^{(0)}. \label{eq:status-constraint}
\end{equation}

In particular, we let $\mbff = \mbff^{(0)} - \mbfx(D,\mbfp)$ be the \emph{residual vector}, which will be dynamically updated when we include new pure strategies into $D$.
The goal is to eventually decrease $\mbff$ to the all-zero vector $\mathbf{0}$.
In such case our algorithm terminates and outputs the mixed strategy $(D,\mbfp)$.
Our algorithm works in iterations.
In each iteration, the algorithm includes $O(n)$ pure strategies into $D$ \footnote{As we may include the same pure strategy into $D$ multiple times in different iterations, $D$ would be a multi-set of pure strategies, in which if a pure strategy appears several times, they are regarded as different strategies and can have different probabilities.}, and guarantees that the inclusion of new strategies increases $|\vmax(\mbff)| + |V_0(\mbff)|$ by at least one.
It can also be verified that throughout the whole algorithm $|\vmax(\mbff)|$ and $|V_0(\mbff)|$ never decrease.
The algorithm terminates when $|\vmax(\mbff)| + |V_0(\mbff)| > n$, in which case we have $\vmax(\mbff)\cap V_0(\mbff) \neq \emptyset$, which implies $\vmax(\mbff) = V_0(\mbff) = V$, i.e., $\mbff$ is an all-zero vector.

\paragraph{Phase A}
A natural idea is to include the pure strategy $\mbfr$ that defends the nodes $\maxtop(\mbff)$, i.e., those with largest $f$ values, into $D$ and give it an appropriate probability satisfying conditions~\eqref{eq:prob-constraint} and~\eqref{eq:status-constraint}.
In particular, suppose $\vmax(\mbff) \subseteq \maxtop(\mbff)$. 
We continuously increase $p(\mbfr)$ (which decreases $f_u$ for all $u\in \maxtop(\mbff)$ at the same rate) until one of the following two events happens
\begin{itemize}
	\item[(a)] the maximum $f$ value of nodes in $\maxtop(\mbff)$ is the same as $\max_{v\notin \maxtop(\mbff)} \{ f_v \}$; or
	\item[(b)] $f_u = 0$ for some $u\in \maxtop(\mbff)$.
\end{itemize}

In Case-(a), we increase $|\vmax(\mbff)|$ by at least one; in Case-(b), we increase $|V_0(\mbff)|$ by at least one without decreasing $|\vmax(\mbff)|$.
In either case, we can finish the iteration with $|\vmax(\mbff)| + |V_0(\mbff)|$ increased by at least one.
The subtle case is when $\vmax(\mbff) \not\subseteq \maxtop(\mbff)$.
In such case, the strategy $\mbfr$ (that defends nodes in $\maxtop(\mbff)$) falls short of defending all nodes in $\vmax(\mbff)$.
As a consequence, we can only have $p(\mbfr) = 0$ because any $p(\mbfr) > 0$ may result in a decrease in $|\vmax(\mbff)|$.
Hence, our algorithm can always enter Phase B.

\paragraph{Phase B}
Observe that $\vmax(\mbff) \not\subseteq \maxtop(\mbff)$ is equivalent to $\sum_{u\in \vmax(\mbff)} \theta_u > R$.
We show that in this case, we can find $O(n)$ pure strategies and associate a probability to each of them such that after including these strategies to $D$, we can decrease $f_u$ for each $u\in \vmax(\mbff)$ by
\begin{equation}
\epsilon := \max_{u\in \vmax(\mbff)} \{f_u\} - \max_{v\notin \vmax(\mbff)} \{f_v\}. \label{eq:epsilon-in-phase-B}
\end{equation}

Lemma~\ref{lemma:epsilon-slice} is the key to find these $O(n)$ pure strategies.

\begin{lemma}\label{lemma:epsilon-slice}
	Given any vector $\mbft \!\in\! \{ 0,\epsilon \}^V$ with $\sum_{u\in \vmax(\mbft)} \theta_u \!>\! R$, we can find $O(|\vmax(\mbft)|)$ pure strategies $T \subseteq \tilOmep(R)$ such that $\| \mbfr \| > R - \theta_\text{max}, \forall \mbfr \in T$.
	Moreover, there exists an integer $c>0$ such that
	\begin{equation*}
	\textstyle \mbft = \frac{\epsilon}{c} \cdot \sum_{\mbfr \in T} \mbfx(\mbfr).
	\end{equation*}		
\end{lemma}

We define $\mbft$ as $t_u = 0$ for all $u\notin \vmax(\mbff)$; $t_u = \epsilon$ for all $u\in \vmax(\mbff)$, where $\epsilon$ is as defined in~\eqref{eq:epsilon-in-phase-B}.
Then by Lemma~\ref{lemma:epsilon-slice}, we can find $O(|\vmax(\mbft)|) = O(n)$ pure strategies $T$ with the above properties.
By giving probability $p(\mbfr) = \epsilon/c$ for each $\mbfr\in T$ and including these strategies into $D$, we can decrease $f_u$ for $u\in \vmax(\mbff)$ by $\epsilon$.
As a consequence, including these new pure strategies increases $|\vmax(\mbff)|$ by at least one.
Hence when this iteration finishes we have $|\vmax(\mbff)| + |V_0(\mbff)|$ increased by at least one.
Observe that in the next iteration we also have $\vmax(\mbff) \not\subseteq \maxtop(\mbff)$.
In other words, the algorithm stays in Phase B until it terminates.

\subsubsection{Proof of Lemma~\ref{lemma:epsilon-slice}}

Let $\mbft\in \{0,\epsilon\}^V$ be the vector given in Lemma~\ref{lemma:epsilon-slice}, and $k = |\vmax(\mbft)|$ be the number of non-zero coordinates.
To prove Lemma~\ref{lemma:epsilon-slice}, we will show that there exists $T\subseteq \tilOmep(R)$ satisfying following conditions:
\begin{itemize}
	\item[(a)] each $\mbfr\in T$ defends only nodes in $\vmax(\mbft)$, i.e., $r_u = \theta_u$ only if $u\in \vmax(\mbft)$;
	\item[(b)] each $\mbfr\in T$ uses total resource $\| \mbfr \| > R-\theta_\text{max}$;
	\item[(c)] there exists an integer $c$ such that for every node $u\in \vmax(\mbft)$, the number of pure strategies in which $u$ is well defended is $| \{ \mbfr\in T : r_u = \theta_u \} | = c$;
	\item[(d)] $|T| \leq k$.
\end{itemize}

Since strategies in $T$ defends only nodes in $\vmax(\mbft)$ and each node in $\vmax(\mbft)$ is defended by the same number of pure strategies, we have $\mbft = \frac{\epsilon}{c}\cdot \sum_{\mbfr\in T} \mbfx(\mbfr)$, as claimed in Lemma~\ref{lemma:epsilon-slice}.
We remark that condition (a) and (b) are relatively easy to satisfy.
The tricky part is to satisfy condition (c) using only $k$ strategies (condition (d)).
We accomplish the mission by proposing the following algorithm.

\smallskip

In the following, we present the ideas for computing $T$.

Let $\{ u_1,\ldots, u_k \}$ be the nodes in $\vmax(\mbft)$, indexed by their IDs.
Suppose $\maxtop(\mbft) = \{ u_1,\ldots,u_i \}$, where $i< k$.
That is, $\sum_{j=1}^i \theta_{u_j} \leq R$ but $\sum_{j=1}^{i+1} \theta_{u_j} > R$.
Then we first include the strategy that defends node in $\maxtop(\mbft)$ and try to find other strategies to defend the remaining nodes $\{ u_{i+1},\ldots,u_k \}$.
Now suppose that $\sum_{j= i+1}^ k \theta_{u_j} \leq R - \theta_\text{max}$.
Then the pure strategy $\mbfr$ that defends only nodes in $\{ u_{i+1},\ldots,u_k \}$ does not satisfy condition (b).
To ensure condition (b) holds, we include nodes $u_1,u_2,\ldots$ into the set of nodes to be defended by $\mbfr$, until we have $\| \mbfr \| > R-\theta_\text{max}$.
In particular, Algorithm~\ref{alg:cyclemt} computes the maximal set of nodes (starting from $u_i$) to be defended, and also returns the end position $j$, i.e., $u_{j-1}$ is defended but $u_j$ is not.
Suppose $(M,j)$ is returned by $\cyclemt(\mbft, i)$.
The main idea is to include the strategy that defends nodes in $M$, and then recursively call $\cyclemt(\mbft, j)$ to compute the next strategy. 
If for some call of $\cyclemt(\mbft, i)$, the returned end position $j = 1$, then we know that the pure strategies we have computed thus far defend all nodes the same number of times.
Unfortunately, we cannot guarantee that this will happen, 	let alone guaranteeing this to happen in $O(k)$ rounds.

Fortunately, we have the following important observation.
Every time when we call the function $\cyclemt(\mbft, i)$, we check whether such a call (with the same parameters $\mbft$ and $i$) has been made before.
If yes, then the set of pure strategies computed since the first call to $\cyclemt(\mbft, i)$ (inclusive) till the second call (exclusive) must have defended all nodes in $\vmax(\mbft)$ the same number of times: the first strategy defends a sequence of nodes starting from node $u_i$, and the last strategy defends a sequence of nodes ending at node $u_{i-1}$.	
In such case we extract this subset of pure strategies and return it as the desired set $T$.
We summarize the steps in Algorithm~\ref{alg:findT}.

\begin{algorithm}[ht]
    \caption{CycleMaxTop($\mbft$, $i$)}
	\label{alg:cyclemt}
	\begin{algorithmic}[1]
	\STATE initialize $M \leftarrow \emptyset$ ; // suppose $\vmax(\mbft) = \{ u_1,\ldots,u_k \}$ \\
	\WHILE{ $\sum_{u\in M} \theta_u \leq R-\theta_\text{max}$ }
		\STATE $M\leftarrow M\cup \{ u_i \}$ 
		\STATE $i\leftarrow 1+ ( i \mod k ) $ 
	\ENDWHILE
	\STATE \textbf{return} ($M$,$i$) \;
	\end{algorithmic}
\end{algorithm}


\begin{algorithm}[ht]
    \small
    \caption{$\findT(\mbft)$}
	\label{alg:findT}
	\textbf{Input}: $\vmax(\mbft) = \{ u_1,\ldots,u_k \}$ \\
    \textbf{Output}: a set of strategy $T$ that defends all nodes in $\vmax(\mbft)$ with same number of times $c$.
	\begin{algorithmic}[1]
	\STATE suppose $\vmax(\mbft) = \{ u_1,\ldots,u_k \}$ 
	\STATE initialize	$T \leftarrow \emptyset$ and $i\leftarrow 1$ 
	\WHILE{\textsf{True}}
		\STATE $(M, i) \leftarrow \cyclemt(\mbft, i)$
		\STATE let $\mbfr$ be defined as $r_u \!=\! \theta_u$ if $u \in M$ and $r_u \!=\! 0$ otherwise
		\IF{$\mbfr\in T$}
			\STATE remove all strategies in $T$ that are included before $\mbfr$ \;
			\STATE pick an arbitrary $u\in \vmax(\mbft)$, and set $c \leftarrow | \{ \mbfr\in T: r_u = \theta_u \} |$ \;
			\STATE \textbf{return} ($T$,$c$) \;
		\ELSE
			\STATE $T\leftarrow T\cup \{ \mbfr \}$ 
		\ENDIF
	\ENDWHILE
	\end{algorithmic}
\end{algorithm}

\begin{proofof}{Lemma~\ref{lemma:epsilon-slice}}
	As argued above, it suffices to show that the computed set of pure strategies meet conditions (a) - (d).
	By the way the strategies are generated condition (a) and (b) are easily satisfied.
	Condition (c) is satisfied because when we observe that function $\cyclemt(\mbft, i)$ is called for the second time with the same parameters, we keep only the strategies they are computed between these two calls.
	As shown above, these pure strategies defend all nodes in $\vmax(\mbft)$ the same number of times.
	Finally, condition (d) is satisfied because we call the function $\cyclemt(\mbft, i)$ only for $i\in [k]$.
	Thus within $k+1$ calls we must have found two calls with the same input parameter $i$, in which case Algorithm~\ref{alg:findT} terminates and outputs at most $k$ strategies (line 7 - 9).	
\end{proofof}

\subsubsection{The Complete Algorithm}

We summarize the steps of our algorithm in Algorithm~\ref{alg:fastmix}, which takes as input the nodes $V$ (where each $u\in V$ has threshold $\theta_u$), a resource bound $R$ and a vector $\mbff^{(0)}$, and outputs a mixed strategy with properties stated in Lemma~\ref{lemma:from-f-to-mixed}.

\begin{algorithm}[ht]
    \scriptsize
	\caption{Compute Mixed Strategy}
	\label{alg:fastmix}
	\textbf{Input}: $V$, $\{ \theta_u \}_{u\in V}$, $R$, and $\mbff^{(0)} \in [0,1]^V$ \\
    \textbf{Output}: The mixed strategy $(D,\mbfp)$ 
    \begin{algorithmic}[1]
    \STATE $D \leftarrow \emptyset$, $\mbff \leftarrow \mbff^{(0)}$ 
	\WHILE{$\|\mbff\| \neq \mathbf{0}$}
		\STATE $M \leftarrow \maxtop(\mbff)$ 
		\IF{$\vmax(\mbff) \subseteq M$}
			\STATE let $\mbfr$ be defined as follows: 
			\STATE $r_u = \theta_u$ if $u \in M$ and $r_u = 0$ otherwise 
			\STATE $D \leftarrow D \cup \{\mbfr\}$ \;
			\STATE $p(\mbfr) \leftarrow \min\{\max_{u \in M }\{f_u\} - \max_{v \notin M }\{f_v\}, \min_{u \in M }\{f_u\}\}$ 
			\STATE update $\mbff \leftarrow \mbff - p(\mbfr) \cdot \mbfx(\mbfr)$ 
		\ELSE
			\STATE $\epsilon \leftarrow \max_{u \in \vmax(\mbff)}\{f_u\} - \max_{v \notin \vmax(\mbff)}\{f_v\}$ 
			\STATE let $\mbft$ be defined as $t_u = \epsilon$ if $u\in \vmax(\mbff)$ and $t_u = 0$ otherwise \;
			\STATE $(T, c) \leftarrow \findT(\mbft)$ 
			\STATE $D \leftarrow D \cup T$ 
			\STATE set $p(\mbfr) \leftarrow \epsilon/c$ for all $\mbfr \in T$ 
			\STATE update $\mbff \leftarrow \mbff - \epsilon/c \cdot \sum_{\mbfr \in T}  \mbfx(\mbfr)$
		\ENDIF
	\ENDWHILE
	\STATE \textbf{return} $(D,\mbfp)$ 
    \end{algorithmic}
\end{algorithm}

Each while loop of Algorithm~\ref{alg:fastmix} correspond to one iteration of our algorithm.
In Particular, line 5 - 8 correspond to Phase A of the algorithm, during which we add one new pure strategy in each iteration; line 10 - 15 correspond to Phase B of the algorithm, which is called only if $\vmax(\mbff) \not\subseteq \maxtop(\mbff)$.
In such case, we let $\mbft$ be defined as we stated in Section~\ref{sssec:overview}, and call the sub-routine $\findT(\mbft)$ (the detailed description of the algorithm is included in the appendix) to compute the set of pure strategies $T\subseteq \tilOmep(R)$ and the constant $c$ as stated in Lemma~\ref{lemma:epsilon-slice}. 
We include the pure strategies in $T$ into $D$, and give each of them probability $\epsilon/c$, which finishes the iteration.

\subsubsection{Analysis}

We first prove the correctness of our algorithm, i.e., the mixed strategy $(D,\mbfp)$ returned by Algorithm~\ref{alg:fastmix} satisfies (1) $\mbfx(D,\mbfp) = \mbff^{(0)}$; (2) $D = O(n^2)$; and (3) $\sum_{\mbfr\in D} p(\mbfr) \leq 1$.

The first condition is easy to show because our algorithm always guarantees that $\mbfx(D,\mbfp) \leq \mbff^{(0)}$, and terminates only if $\mbff = \mbff^{(0)} - \mbfx(D,\mbfp)$ is an all-zero vector.
Following the arguments we have presented, in each iteration of Phase A we include one pure strategy into $D$; in each iteration of Phase B we include $O(n)$ pure strategies into $D$ (by Lemma~\ref{lemma:epsilon-slice}).
Since each iteration increases $|\vmax(\mbff)| + |V_0(\mbff)|$ by at least one and our algorithm terminates when $|\vmax(\mbff)| + |V_0(\mbff)| > n$ (in which case we have $\mbff = \mathbf{0}$), we conclude that there are at most $n$ iterations.
Hence $|D| = O(n^2)$.

Next we show that $\sum_{\mbfr\in D} p(\mbfr) \leq 1$.
We analyze total probability in two cases, depending on whether algorithm ever enters Phase B.

Let $u\in \vmax(\mbff^{(0)})$ be an arbitrary node with maximum $f_u$ in $\mbff^{(0)}$.
Throughout the whole algorithm, we can guarantee $u\in \vmax(\mbff)$ for any $\mbff$, because we never decrease $f_u$ to a value that is lower than the second largest $f$ value.
Hence if Algorithm~\ref{alg:fastmix} never enters Phase B, then we have
\begin{equation*}
\textstyle \sum_{\mbfr\in D} p(\mbfr) = f_u \leq 1.
\end{equation*}

Now suppose that Algorithm~\ref{alg:fastmix} terminates at Phase B.
The important observation here is that in such case we have $\sum_{u\in M} \theta_u > R - \theta_\text{max}$ for every $M$ that is returned by $\maxtop(\mbff)$ in line 3,
because $\sum_{u\in M} \theta_u \leq R - \theta_\text{max}$ happens only if $\maxtop(\mbff) = V\setminus V_0(\mbff)$, in which case the algorithm never enters Phase B.
Consequently for each $\mbfr\in D$ we have $\| \mbfr \| > R-\theta_\text{max}$.
Recall that $\mbff^{(0)}$ is defined such that for some $\tilmbfr^* \in \Omega_p(R - \theta_\text{max})$, $f^{(0)}_u = \min\{ \tilde{r}^*_u / \theta_u, 1 \}$ for all $u\in V$.
Also recall that for each $u\in V$ we have
\begin{equation*}
\sum_{\mbfr \in D} (p(\mbfr)\cdot r_u)
= \theta_u \cdot x_u(D, \mbfp) = \theta_u \cdot f^{(0)}_u.
\end{equation*}
It follows that
\begin{equation}
\sum_{u\in V} \sum_{\mbfr \in D} (p(\mbfr)\cdot r_u)
= \sum_{u\in V} (\theta_u\cdot f^{(0)}_u) \leq \sum_{u\in V} \tilde{r}^*_u \leq R - \theta_\text{max}.
\label{eq:upper-bound-of-x-norm}
\end{equation}

On the other hand, we have
\begin{equation}
\sum_{u\in V}\sum_{\mbfr \in D} (p(\mbfr)\cdot r_u)
= \sum_{\mbfr \in D} (p(\mbfr)\cdot \| \mbfr \|)
> (R - \theta_\text{max})\cdot \sum_{\mbfr \in D} p(\mbfr).
\label{eq:lower-bound-of-x-norm}
\end{equation}

Combining~\eqref{eq:upper-bound-of-x-norm} and~\eqref{eq:lower-bound-of-x-norm}, we have $\sum_{\mbfr \in D} p(\mbfr) < 1$.

\paragraph{Complexity.}
Now we analyze the complexity of Algorithm~\ref{alg:fastmix}.
It is easy to check that $\vmax(\mbff)$, $V_0(\mbff)$, $\maxtop(\mbff)$ and $\cyclemt(\mbft, i)$ can be computed in $O(n\log n)$ time.
From the proof of Lemma~\ref{lemma:epsilon-slice}, we know that Algorithm~\ref{alg:findT} executes in $O(n)$ rounds.
Thus each call to $\findT(\mbft)$ finishes in $O(n^2 \log n)$ time.
Finally, since there are $O(n)$ iterations, and in each iteration $\findT(\mbft)$ is called at most once, the total complexity of Algorithm~\ref{alg:fastmix} is $O(n^3\log n)$.

\subsection{Mixed Strategy with Resource Sharing}\label{ssec:hardness-sharing}
	
	As we have shown above, in the isolated model we can give a strong upper bound $\opt_m(R)$ by $\opt_f(R-\theta_\text{max})$.
	It would be natural to ask whether similar upper bounds hold under the non-isolated model, i.e., when defending resource can be shared between neighboring nodes.
	Unfortunately, we show that when resource sharing is allowed, we do not have such guarantees, even if we allow the mixed strategy to use several times more resource than the fractional strategy.
	
	\begin{lemma}\label{lemma:hardness-shared}
		For any constant $\beta > 1$, there exists an instance for which $\opt_m(\beta \cdot R) >  \opt_f(R)$.
	\end{lemma}
	\begin{proof}
		Consider a complete bipartite graph $G(U\cup V, E)$, where $|U| = 2 \beta \cdot R, |V| = 4 \beta^2 \cdot R$ and all edges $(u,v)\in E = U\times V$ have the same weight $w_{uv} = 1/|U|$.
		Let $\theta_u= \alpha_u = 1$ for all $u\in U\cup V$.
		
		Observe that there exists a fractional strategy (using total resource $R$) that allocates $1/(2\beta)$ resource to each of the nodes in $U$, under which every node in $U$ and $V$ has defending power $1/(2\beta)$.
		Therefore, the defending result of this fractional strategy is $1-1/(2\beta)$, which implies that $\opt_f(R) \leq 1-1/(2\beta)$.
		Next we show that for every $\mbf{r} \in \Omega_p(\beta\cdot R)$, the number of well defended nodes is at most $2  \beta \cdot R$.
		%
			Suppose otherwise, there must exist a well defended node $u$ with $r_u < \frac{\beta\cdot R}{2\beta\cdot R} = 0.5$.
			Hence
			\begin{align*}
			 \pi_u(\mbf{r}) & \textstyle = r_u + \sum_{v\in N(u)} w_{uv}\cdot r_v  \\
			& \textstyle \leq r_u + \frac{1}{2\beta\cdot R} \cdot (\beta\cdot R - r_u) < \frac{1}{2} + \frac{1}{2} = 1.
			\end{align*}
			However, since $u$ is well defended, we must have $\pi_u(\mbf{r}) \geq 1$, which is a contradiction.
		
		Hence for any pure strategy $\mbf{r}\in \Omega_p(\beta\cdot R)$, we have $\| \mbf{x}(\mbf{r}) \| \leq 2 \beta \cdot R$.
		Consequently for any mixed strategy $(D,\mbf{p}) \in \Omega_m(\beta\cdot R)$, we have $\| \mbf{x}(D,\mbf{p}) \| \leq 2 \beta \cdot R$ (because $\mbf{x}(D,\mbf{p})$ is a linear combination of defending statuses of pure strategies).
		Hence there must exist a node $u$ for which 
		\begin{equation*}
		\textstyle	{x}_u(D,\mbf{p}) \leq \frac{2 \beta \cdot R}{|U| + |V|} = \frac{2 \beta \cdot R}{2 \beta \cdot R + 4 \beta^2 \cdot R} < \frac{1}{2 \beta}.
		\end{equation*}
			
		Therefore, $\opt_m(\beta\cdot R) \!>\! 1-1/(2\beta) \!\geq\! \opt_f(R)$.
	\end{proof}

    \section{Small Support Mixed Strategies}\label{sec:algorithm}
    
    \newcommand{\ProbLP}{\textsf{ProbLP}}
    \newcommand{\FindR}{\textsf{FindR}}

	So far, we evaluate the quality of a mixed strategy only by its defending result without considering its support size $|D|$.
	Intuitively, the larger support a mixed strategy has, the more likely the strategy can balance the defending status among all nodes.
	However, in practice, it is usually preferable to have mixed strategies $(D,\mbfp)$ with a small $D$ for efficiency purpose.
	In this section, we study the computation of mixed strategies that have good defending results and small support.
	In particular, we propose the \algName\space algorithm that computes mixed strategies with an upper bound on the support size, but also have good defending results.
	By Theorem~\ref{thy:hardness}, we know that computing the optimal mixed strategy is \textsf{NP}-hard.
	Moreover, by the reduction we can see that even computing the optimal mixed strategy with $|D| = 2$ is \textsf{NP}-hard.
	However, we have the following very helpful observations.
	We show that deciding if a set of nodes can be defended simultaneously using one pure strategy is polynomial-time solvable.
	Throughout this section we fix $G(V,E)$ to be the graph instance and $R$ to be the total resource.

	\begin{lemma}\label{lemma:from-S-to-r}
		Given a set of nodes $S\subseteq V$, deciding if there exists $\mbfr\in \Omega_{p}$ with $x_u(\mbfr) = 1$ for all $u\in S$ is polynomial-time solvable.
		Moreover, if they exist, we can compute one in polynomial time.
	\end{lemma}
	\begin{proof}
		We can reduce the problem of defending all nodes in $S$ with one pure strategy (using total resource $R$) to solving the following feasibility LP. 
		In particular, we introduce the variable $r_u$ to denote the resource allocated to node $u$.
		We introduce the constraints that total resource used is at most $R$, and that each node $u\in S$ has defending power at least $\theta_u$.
		\begin{align*}
		\text{minimize }\qquad & \qquad 0 \\
		\text{subject to }\qquad & \textstyle \sum_{u\in V} r_u \leq R, \\
		& \textstyle r_u + \sum_{v\in N(u)} w_{vu} \cdot r_v \geq \theta_u, \qquad \forall u \in S \\
		& \textstyle r_u \geq 0, \qquad \forall u \in V.
		\end{align*}
		
		If the above LP is infeasible then there does not exist a pure strategy that can defend all nodes in $S$; otherwise any feasible solution to the LP is the desired pure strategy.	
	\end{proof}
	
	While computing the optimal mixed strategy is \textsf{NP}-hard, we show that for a small set of pure strategies $D$, computing the optimal mixed strategy with support $D$ can be solved in polynomial time.
	
	\begin{lemma}\label{lemma:optimal-for-fixed-D}
		Given a set of pure strategies $D\subseteq \Omega_{p}$, the optimal mixed strategy $(D, \mbfp^*)$ with support $D$ can be computed in time polynomial in $|D|$ and $n$.
	\end{lemma}
	\begin{proof}
		Since $D$ is fixed, the problem is to decide the probability $p(\mbfr)$ for each $\mbfr\in D$, such that the defending result is as small as possible.
		We first compute the defending status $\mbfx(\mbfr)$ for each $\mbfr\in D$.
		Then we transform this problem into an LP, in which the probabilities $\{ p(\mbfr) \}_{\mbfr\in D}$ and $L$ are variables.
		\begin{align*}
		\text{minimize }\qquad & \qquad L \\
		\text{subject to }\qquad & \textstyle \sum_{\mbfr\in D} p(\mbfr) = 1, \\
		& \textstyle (1 - \sum_{\mbfr\in D} p(\mbfr)\cdot x_u(\mbfr) ) \cdot \alpha_u \leq L,   \; \forall u \in V\\
		& \textstyle p(\mbfr) \geq 0, \qquad \forall \mbfr\in D.
		\end{align*}
		
		It can be verified that the optimal solution to the above LP corresponds to the mixed strategy with support $D$ that has minimum loss.
		The first and third sets of constraints guarantee that $\{ p(\mbfr) \}_{\mbfr\in D}$ is a feasible probability distribution over $D$.
		The second set of constraints guarantee that the final defending result is minimum.
	\end{proof}
	
	\subsection{The {\algName} Algorithm}
	
	Following the above observations, we propose the following local-search based algorithm that progressively and efficiently computes a mixed strategy with small support and good defending result. 
	Our algorithm takes as input an iteration bound $d$, terminates after $d$ search steps and outputs a mixed strategy $(D,\mbfp)$ with $|D|\leq d$.
	For convenience of notation we use $L_m(u)$ to denote $L_m(u, (D,\mbfp))$, when the mixed strategy $(D,\mbfp)$ is clear from the context.
	
	Intuitively speaking, our algorithm starts from a mixed strategy $(D,\mbfp)$ and tries to include a new pure strategy $\mbfr$ into $D$, so that the optimal mixed strategy with support $D\cup \{ \mbfr \}$ is likely to achieve a better defending result.
	As shown in Lemma~\ref{lemma:optimal-for-fixed-D}, as long as $|D|$ is small, computing the optimal mixed strategy with support $D$ can be done efficiently by solving an LP.
	We denote this sub-routine by $\ProbLP(D)$.
	Our main idea is to add the new strategy to patch the poorly defended nodes up based on their current losses.
	Borrowing some ideas from the proof of Lemma~\ref{lemma:from-f-to-mixed}, we compute the maximal set of nodes $M$ with largest losses under the current mixed strategy $(D,\mbfp)$, and use Lemma~\ref{lemma:from-S-to-r} to compute a new pure strategy that defends these nodes.
 	As we will show in the next section, as long as the maximum loss of nodes in $M$ is larger than that of nodes not in $M$, our algorithm can always make progress in decreasing the defending result.
 	Otherwise we randomly permute the nodes in $V$ and try to include a random new pure strategy into $D$.
	We introduce the \FindR($L_m$) subroutine for the computation of the new pure strategy, for a given loss vector $L_m$.
	Note that if it fails to compute a new pure strategy, an all-zero vector will be returned.
	We summarize the main steps of the {\algName} algorithm in Algorithm~\ref{alg:patching}.
	Initially we set the strategy set $D$ to be a singleton containing only the optimal pure strategy and the algorithm terminates after $d$ iterations.	
	

	
	\begin{algorithm}[tb]
        \caption{\algName}
        \label{alg:patching}
        \textbf{Input}: the number of iterations $d$ and optimal pure strategy $\mbfr^*$\\
        \textbf{Output}: mixed strategy set $\mstg$ 
		\begin{algorithmic}[1] 
		\FOR{$i = 2,3,\ldots, d$}
			\STATE $\mbfp \gets  \ProbLP(D)$ 
			\STATE compute the loss vector $L_m$ of mixed strategy $(D,\mbfp)$ 
			\STATE $\mbfr \gets \FindR(L_m)$ 
			\IF{$\| \mbfr \|\neq 0$}
				\STATE $D\gets D\cup \{ \mbfr \}$ 
			\ENDIF
		\ENDFOR
		\STATE $\mbfp \gets  \ProbLP(D)$\;
		\STATE \textbf{return} $\mstg$\\
		\end{algorithmic}
	\end{algorithm} 
	
	
	\begin{algorithm}[tb]
		\caption{\FindR}
		\label{alg:finfr}
		\textbf{Input}: the loss vector $L_m$ with current strategy set $D$\\
        \textbf{Output}: new pure strategy $\mbfr$
		\begin{algorithmic}[1]
		\STATE let $M \gets \SMaxTop(L_m)$ 
		\IF{$\exists\ \mbfr \in D: x_u(\mbfr) = 1$ for all $u\in M$}
			\STATE replace $L_m$ with a random vector in $[0,1]^V$ 
			\STATE let $M \gets \SMaxTop(L_m)$ 
			\IF{$\exists\ \mbfr \in D: x_u(\mbfr) = 1$ for all $u\in M$}
				\STATE \textbf{return} $\{0\}^V$ 
			\ENDIF
		\ENDIF
		\STATE let $\mbfr$ be the pure strategy that defends all nodes in $M$ 
		\STATE \textbf{return} $\mbfr$ 
		
        \end{algorithmic}
	\end{algorithm}

	
	Next we introduce the details of the sub-routine \FindR.
	As discussed, given the loss vector $L_m$, the idea is to first locate the nodes with large losses and then generate a new pure strategy that enhances the defending statuses of these poorly defended nodes. 
	We thus use similar ideas as in the proof of Lemma~\ref{lemma:from-f-to-mixed} to compute the maximal set of nodes to be defended.
	However, since resource sharing is considered, the procedure is slightly more complicated.
	
	Given any integer $k>0$, we can identify the Top-$k$ nodes $S$ with maximal losses in $L_m$, and check whether it is possible to defend all nodes in $S$ by solving an LP (see Lemma~\ref{lemma:from-S-to-r}).
	Using a binary search on $k$ we can identify the maximum $k$ for which the corresponding set of nodes $S$ can be defended.
	Let $\SMaxTop(L_m)$ be these nodes.
%
	The sub-routine $\FindR(L_m)$ first computes $M\gets \SMaxTop(L_m)$ and tries to include the pure strategy $\mbfr$ that defends all nodes in $M$.
	If there already exists a strategy in $D$ that defends all nodes in $M$, then it is unnecessary to include $\mbfr$ because its inclusion will not help in decreasing the defending result.
	In such case we do a random permutation on $V$ (by replacing $L_m$ with a random vector in $[0,1]^V$), and compute another pure strategy.
	As mentioned, if we fail to find a new pure strategy after the random permutation, then the sub-routine returns the trivial defending strategy $\{0\}^V$.
	We summarize the steps of $\FindR(L_m)$ in Algorithm~\ref{alg:finfr}.

	\paragraph{Complexity.}
	Observe that every call to $\SMaxTop$ involves $O(\log n)$ computations of some feasibility LPs with $O(n)$ variables.
	Therefore, the total complexity of the {\FindR} algorithm is bounded by $O(\log n)$ computations of LPs.
	Note that the complexity of each iteration of the {\algName} algorithm is dominated by the {\FindR} sub-routine (recall that {\ProbLP} can be done by solving one LP).
	As a consequence, the total complexity of {\algName} is bounded by $O(d \log n)$ computations of LPs, where $d$ is the number of iterations.

	\subsection{Effectiveness}
	
	As we will show in our experiments (Section~\ref{sec:experiments}), the \algName\space algorithm achieves close-to-optimal defending results on several large datasets.
	In this section, we theoretically analyze the algorithm and formalize the condition under which our algorithm is guaranteed to make progress in terms of decreasing the defending result.
	Our analysis also sheds lights into why random permutation could help in improving the performance of the algorithm.
	
	Consider any iteration of the \algName\space algorithm.
	Suppose $(D,\mbfp)$ is the current mixed strategy and $L_m$ is the loss vector.
	In line 5 of Algorithm~\ref{alg:patching}, we call the sub-routine \FindR.
	In the sub-routine we compute the maximal set of nodes that can be defended $M \gets \SMaxTop(L_m)$ (line 1 of Algorithm~\ref{alg:finfr}).
	Lemma~\ref{lemma:progress-patching} states that as long as $M$ contains all nodes with maximum loss (in which case $\Delta L > 0$), our algorithm can always make progress in decreasing the defending result.
	
	\begin{lemma} \label{lemma:progress-patching}
		 Let $\mbfr$ be the pure strategy that defends all nodes in $M$.
		 Including $\mbfr$ into $D$ decreases the defending result of the current mixed strategy $(D,\mbfp)$ by at least $\frac{\Delta L}{ \Delta L + \alpha_\text{max}} \cdot L_m(D, \mbfp)$,
		where
		\begin{small}
		\begin{equation*}
    		\alpha_\text{max} = \max_{u\in V} \{ \alpha_u \} \quad \text{and} \quad \Delta L = \max_{u\in M} \{ L_m(u) \} - \max_{u\notin M} \{ L_m(u) \}.
		\end{equation*}
		\end{small}
	\end{lemma}
	\begin{proof}
		Let $\epsilon = \frac{\Delta L}{ \Delta L + \alpha_\text{max}}$ and $D' = D\cup \{ \mbfr \}$.
		We show that there exists a mixed strategy with support $D'$ that achieves defending result $(1-\epsilon)\cdot L_m(D,\mbfp)$.
		Specifically, we define the mixed strategy $(D', \mbfp')$ as follows.
		Let $p'(\mbfr) = \epsilon$; for each $\mbfr' \in D$, let $p'(\mbfr') = (1-\epsilon)\cdot p(\mbfr')$.
		Since $\| \mbfp \| = 1$, we have $\| \mbfp' \| = \epsilon + (1-\epsilon)\cdot \| \mbfp \| = 1$.
		Thus $(D', \mbfp')$ is a feasible mixed strategy.
		For each $v\notin M$, since $v$ is not defended by the new strategy $\mbfr$, we have
		\begin{small}
		\begin{align*}
		&\textstyle L_m(v, (D', \mbfp')) = \left(1 - \sum_{\mbfr \in D'} p'(\mbfr) \cdot x_v(\mbfr)\right) \cdot \alpha_v \\
		= &\textstyle \left(1 - \sum_{\mbfr \in D} p'(\mbfr) \!\cdot\! x_v(\mbfr)\right) \cdot \alpha_v \\
		= & \textstyle \left(1 - (1 - \epsilon) \!\cdot\! \sum_{\mbfr \in D} p(\mbfr) \cdot x_v(\mbfr)\right) \cdot \alpha_v  \\
		= & (1 - \epsilon) \cdot L_m(v, (D,\mbfp)) + \epsilon \cdot \alpha_v \\
		\leq & (1 - \epsilon) \cdot ( L_m(D,\mbfp) - \Delta L ) + \epsilon \cdot \alpha_\text{max} = (1 - \epsilon) \cdot L_m(D,\mbfp).
		\end{align*}
		\end{small}
		
		For each $u\in M$, we have
		\begin{small}
		\begin{align*}
		L_m(u, (D',\mbfp')) &\textstyle = \left(1 - \sum_{\mbfr \in D'} p'(\mbfr) \cdot x_u(\mbfr)\right) \cdot \alpha_u \\
		&\textstyle = \left(1 - (1 - \epsilon) \!\cdot\! \sum_{\mbfr \in D} p(\mbfr) \cdot x_u(\mbfr) - \epsilon \cdot 1\right) \cdot \alpha_u \\
		& = (1 - \epsilon) \!\cdot\! L_m(u, (D,\mbfp)) \!\leq\! (1 - \epsilon) \cdot L_m(D,\mbfp).
		\end{align*}
		\end{small}
		
		Hence we have $L_m(D', \mbfp') \leq (1-\epsilon)\cdot L_m(D,\mbfp)$.		
		So given $D' = D\cup \{ \mbfr \}$, when our algorithm computes the optimal mixed strategy with support $D'$, its defending result must be at most $(1-\epsilon)\!\cdot\! L_m(D,\mbfp)$, as claimed by the lemma.
	\end{proof}
	
	By Lemma~\ref{lemma:progress-patching}, we can see that as long as all nodes with maximum loss can be defended by one pure strategy, the \algName\space algorithm can always decrease the defending result.
	When the nodes with maximum loss are too many and no pure strategy can defend them all, we randomly permute the nodes to compute a random pure strategy to be included in $D$.
	As we observed from our empirical study, such random permutations are crucial as otherwise the algorithm may get stuck in the early stage during the execution.

	\section{Experimental Evaluation} \label{sec:experiments}
	
	In this section we perform the experimental evaluation of our algorithms on several real-world graph datasets whose sizes range from 1000 nodes to 260k nodes (see Table~\ref{table:datasets}).
	All the datasets are downloaded from SNAP by Stanford~\citep{snapnets}. 
	Unless otherwise specified, we set the parameters of instances as follows.\footnote{We remark that for different settings of the parameters, e.g., wider ranges for the values and thresholds, or smaller values of $R$, the experimental results are very similar.} 
	For each of these datasets, we set the value $\alpha_u$ of each node $u$ to be an independent random integer chosen uniformly at random from $[1,9]$.	
	We set the threshold $\theta_u$ of each node $u$ to be an independent random real number chosen uniformly at random from $[1,10]$.
	We set the weight $w_{uv}$ of each edge $(u,v)\in E$ to be an independent random real number chosen from $[0,1]$.
	We set the total resource $R = 0.2\cdot \sum_{u\in V} \theta_u$.
	
	\begin{table}[ht]
		\caption{Number of nodes and edges of the datasets}
		\label{table:datasets}
		\centering
		\scriptsize
		\setlength\tabcolsep{4pt}
		\begin{tabular}{ c | c | c | c | c | c | c}
			\toprule
			Dataset & Email-S & Facebook & Ca-AstroPh & Email-L & Twitter & Amazon  \\
			\midrule
			\# Nodes & 1,005 & 4,039 & 18,772 & 36,692 & 81,306 & 262,111  \\
			\# Edges & 27,551 & 88,234 & 198,110 & 367,662 & 1,768,149 & 1,234,877  \\
			\bottomrule
		\end{tabular}
	\end{table}
	
	
	In the experiments we mainly evaluate the effectiveness of the {\algName} algorithm.
	Additionally we test and report the mixed strategies we stated in Section~\ref{ssec:upperbound}, and compare their defending results and support sizes with that of the mixed strategies returned by {\algName}. 
	As we have shown in Section~\ref{sec:computation}, for the same problem instance we always have $\opt_m \geq \opt_f$.
	Thus in our experiments we mainly use $\textsf{OPT}_f$ as the baseline to evaluate the performance of the mixed strategies.
	
	\paragraph{Experiment Environment.}
	We perform our experiments on an AWS Ubuntu 18.04 machine with 32 threads and 128GB RAM without GPU. We use Gurobi optimizer as our solver for the LPs.

	\subsection{Uniform Threshold in the Isolated Model}\label{ssec:Noshare}
	
\begin{figure}[htbp]
		\centering
		\includegraphics[width=\linewidth]{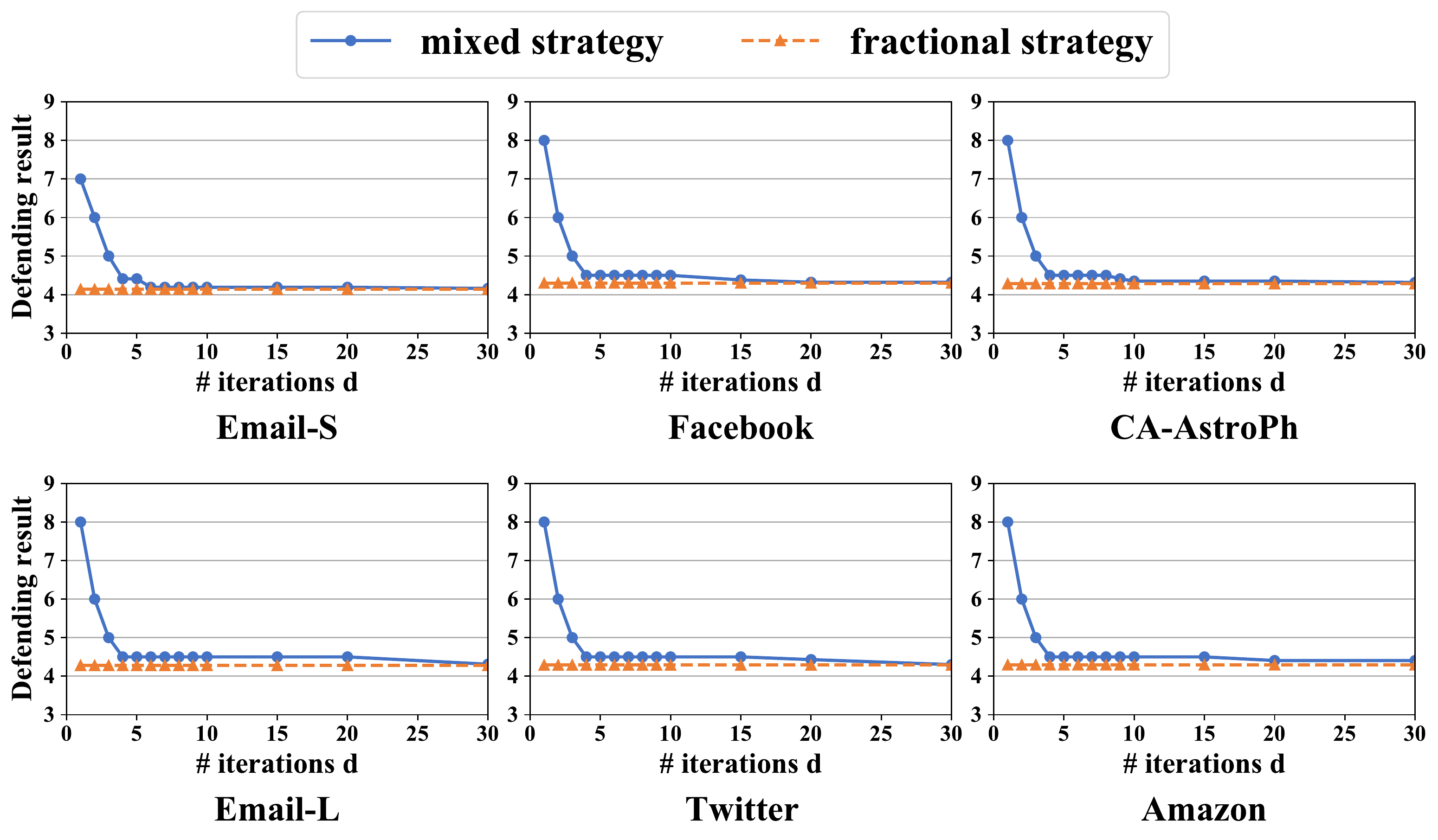}
		\caption{Defending Results of Mixed Strategies by {\algName} in Uniform Threshold Isolated Model.}
	\label{fig:nosharing}
    \end{figure}

	We first consider the most basic setting with uniform threshold and without resource sharing.
	The same setting was also considered in~\citep{atal/KiekintveldJTPOT09,aaai/KorzhykCP10}.
	That is, we set $\theta_u = 1$ for all nodes $u\in V$ and $w_{uv} = 0$ for all edges $(u,v)\in E$, in this subsection.
	Under this setting it can be shown that $\opt_m = \opt_f$ for all instances of the problem~\citep{atal/KiekintveldJTPOT09,aaai/KorzhykCP10}.
	Thus we measure the effectiveness of the {\algName} algorithm by comparing the defending result of the returned mixed strategy with $\opt_f$.
	For each dataset, we report the defending result of the mixed strategy returned by \algName$(d)$, for increasing values of $d \in [1,30]$.
    Note that for $d= 1$, the mixed strategy uses the optimal pure strategy $\mbfr^*$ with probability $1$.
	For general $d$, the returned mixed strategy uses $|D|\leq d$ pure strategies.
	The results are presented as Figure~\ref{fig:nosharing}.
	
	From Figure \ref{fig:nosharing}, we observe that the defending result  rapidly decreases in the first few iterations of the algorithm, and gradually converges to the optimal defending result $\opt_f$.
	Specifically, within the first $5$ iterations, the defending result of the mixed strategy is already within a $5\%$ difference with the optimal one in most datasets.
	After $30$ iterations, the defending result is almost identical to the optimal one in all datasets.
	The experiment demonstrates the effectiveness of our algorithm on computing mixed strategies that have small support sizes and are close-to-optimal.

	\subsection{General Thresholds in the Isolated Model}\label{ssec: general_theta_isolated}
	
\begin{figure}[htb]
		\centering
		\includegraphics[width=\linewidth]{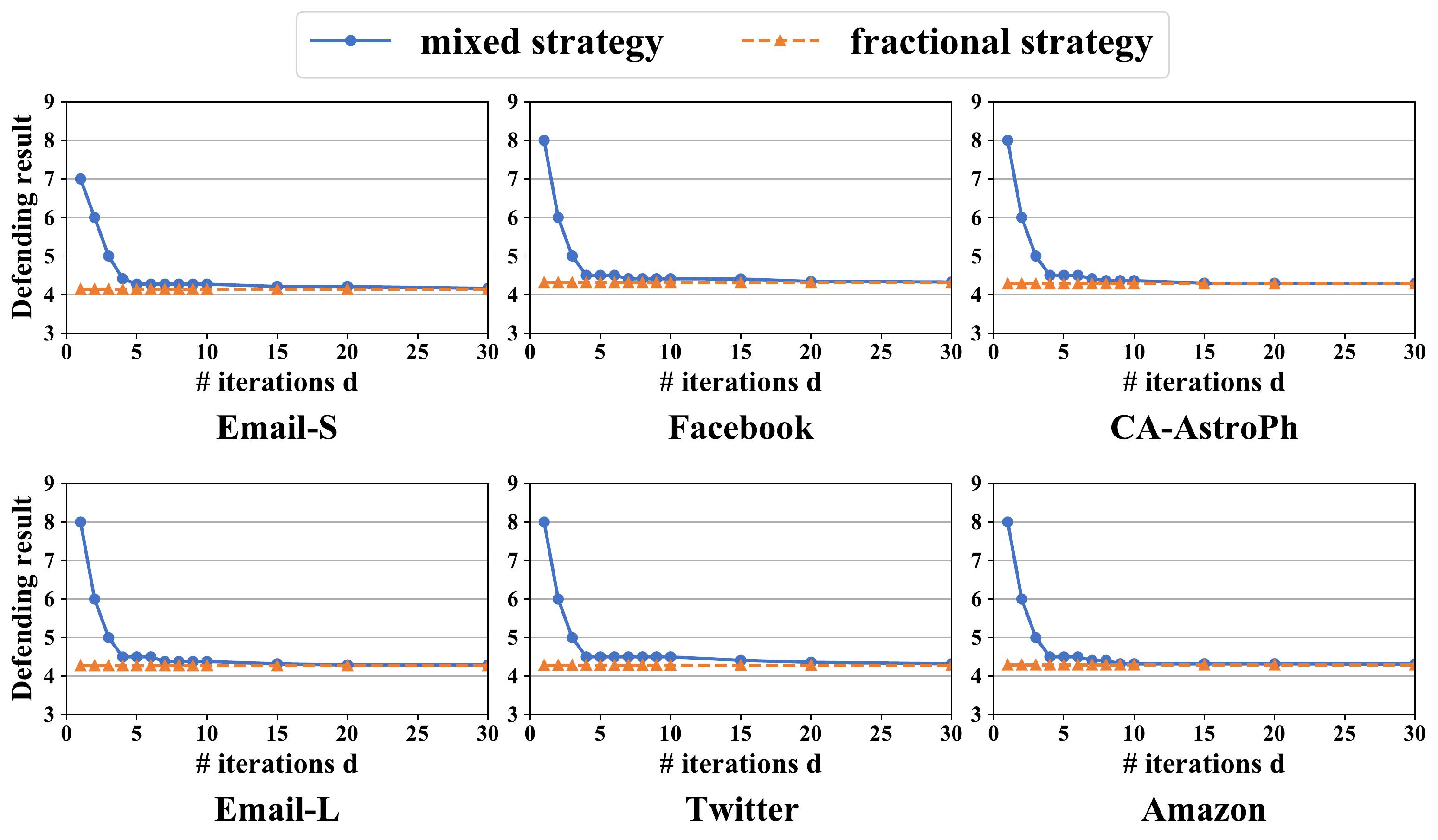}
		\caption{Defending Results of Mixed Strategies by {\algName} in General Threshold Isolated Model.}
		\label{fig:theta_general}
    \end{figure}

	Next we consider the more general setting with non-uniform thresholds, e.g., $\theta_u\in [1,10]$ is chosen uniformly at random for each node $u\in V$.
	As we have shown in Theorem~\ref{thy:hardness}, computing the optimal mixed strategy in this case is \textsf{NP}-hard.
	On the other hand, we have shown in Section~\ref{ssec:upperbound} that for any instance we always have
	$\opt_f(R) \leq \opt_m(R) \leq \opt_f(R-\theta_\text{max})$, where the maximum threshold $\theta_\text{max} \leq 10$ in our experiments.
	The experimental results are reported in Figure \ref{fig:theta_general}.
	
	As we can observe from Figure \ref{fig:theta_general}, the result is very similar to the uniform threshold case we have considered in the previous experiments.
	This also confirms our theoretical analysis in Section~\ref{ssec:upperbound}: when $R$ is sufficiently large (compared with $\theta_\text{max}$), the optimal mixed strategy should have defending result close to $\opt_f(R)$.
	Furthermore, the experiment demonstrates that even for general thresholds, {\algName} returns mixed strategies that are close-to-optimal, and uses very few pure strategies.
	

    Recall that in Section~\ref{ssec:upperbound} we show that there exists a mixed strategy $(D,\mbfp)\in \Omega_{m}(R)$ whose defending result $L_m(D,\mbfp) = \opt_f(R-\theta_\text{max})$.
	In our experiments, we implement the algorithm and report the defending result and the support size to verify its correctness.
	The results are presented in Table~\ref{table:losscompare}, where $(D,\mbfp)$ is the mixed strategy our algorithm computes, and $(D,\mbfp^*)$ is the optimal mixed strategy with support $D$ (which can be computed by solving an LP, as we have shown in Lemma~\ref{lemma:optimal-for-fixed-D}).
	Note that due to long computation time, we did not finish the computing of $L_m(D,\mbfp^*)$ for the Twitter and Amazon datasets (too many variables).
	In the table, we also compare them with the mixed strategies returned by {\algName}$(d)$ with $d \in \{ 5,30 \}$.
	Consistent with our theoretical analysis, for all datasets we have $\opt_f(R-\theta_\text{max}) = L_m(D,\mbfp) \geq L_m(D,\mbfp^*) \geq \opt_f(R)$.
	From the above results we can see that compared to $(D,\mbfp)$, the {\algName} algorithm is able to compute mixed strategies with very small support, e.g., $30$ vs. $2500+$ for large datasets, while guaranteeing a defending result that is very close.
	
	\begin{table}[ht]
		\caption{Different Mixed Strategies in Different Datasets}
		\label{table:losscompare}
		\centering
		\scriptsize
		\setlength\tabcolsep{2pt}
		\begin{tabular}{ c | c | c | c | c | c | c}
				\toprule
				& \ Email-S & Facebook & Ca-AstroPh & Email-L & Twitter & Amazon \\
				\midrule	        	$\textsf{OPT}_f(R-\theta_\text{max})$ & 4.161 & 4.32 & 4.281 & 4.273 & 4.285 & 4.293 \\
				$L_m(D, \mbfp)$ & 4.161 & 4.32 & 4.281 & 4.273 & 4.285 & 4.293 \\
				$L_m(D, \mbf{p^*})$ & 4.147 & 4.316 & 4.28 & 4.273 & -- & -- \\ 
				$|D|$ & 268 & 671 & 1900 & 2592 & 6052 & 9957 \\
				$
				\textsf{OPT}_f(R)$ & 4.139 & 4.314 & 4.28 & 4.273 & 4.285 & 4.293 \\
				\midrule
				Patching(5) & 4.41 & 4.5 & 4.5 & 4.5 & 4.5 & 4.5\\
				Patching(30) & 4.161 & 4.326 & 4.29 & 4.291 & 4.324 & 4.319 \\
				\bottomrule
		\end{tabular}
	\end{table}
	
	We also compare the time to compute $L_m(D,\mbfp)$ and the running time (in seconds) of {\algName} in Table~\ref{table:patching-time}.
	As we can observe from the table, compared to the computation of $L_m(D,\mbfp)$, the running time of {\algName} is less sensitive to the size of the network.
	Consequently for large datasets the {\algName} algorithm runs several times faster than computing $L_m(D,\mbfp)$.
	In conclusion, the {\algName} algorithm computes mixed strategies that use way fewer pure strategies than $(D,\mbfp)$ while having similar defending results. Furthermore, its running time is also much smaller in large datasets.
	
		\begin{table}[htb]
			\caption{Running Time Comparison (in seconds)}
		    \label{table:patching-time}
			\centering
			\scriptsize
			\setlength\tabcolsep{3pt}
			\begin{tabular}{c | c | c | c | c | c | c}
				\toprule
				& \ Email-S & Facebook & Ca-AstroPh & Email-L & Twitter & Amazon \\
				\midrule
				$L_m(D, \mbfp)$ & 0.2 & 3.2 & 57 & 214 & 1027 & 9370 \\
			    Patching(5) & 0.3 & 0.8 & 3.3 & 6.5 & 14 & 45 \\
				Patching(30) & 2.3 & 8.2 & 37 & 73 & 165 & 523\\
				\bottomrule
			\end{tabular}
		\end{table}

	\subsection{General Thresholds with Resource Sharing}
	\label{ssec:sharing}
	
\begin{figure}[htb]
		\centering
		\includegraphics[width=\linewidth]{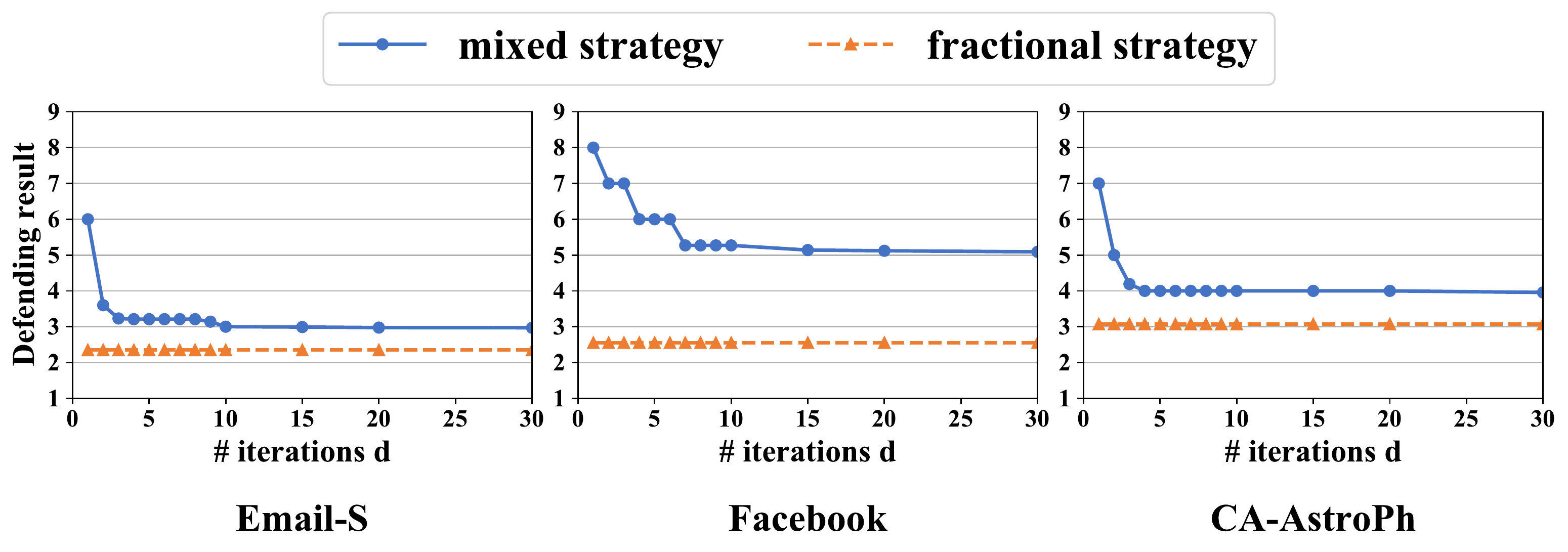}
		\caption{{\algName} in the Non-isolated Model}
		\label{fig:sharing}
    \end{figure}

	Finally, we evaluate the performance of the {\algName} algorithm on the network defending problem with resource sharing.
	In contrast to the isolated model, with resource sharing we can no longer guarantee $\opt_m \approx \opt_f$, even if $R$ is sufficiently large (see Section~\ref{ssec:hardness-sharing} for the hard instance).
	In other words, the lower bound $\opt_f$ we compare our mixed strategy with can possibly be much smaller than the optimal defending result $\opt_m$ of mixed strategies.
	Moreover, with resource sharing we must set $R$ to be smaller, as otherwise, e.g., $R = 0.2\cdot \sum_{u\in V} \theta_u$, the defending result is $0$.
	For this reason, we set $R = 0.1 \cdot \sum_{u \in V}{\theta_u}$ for the Email-EU and CA-AstroPh datasets, and $R = 0.015 \cdot \sum_{u\in V}{\theta_u}$ for the Facebook dataset, since slightly larger values of $R$ give $\opt_p = 0$.
	The experimental results are reported as Figure \ref{fig:sharing}.
	As discussed in Section~\ref{sec:algorithm}, computing mixed strategies for the non-isolated model involves solving LPs with $\Theta(n)$ variables, which can be quite time consuming.
	Thus we only manage to run the experiments on the three small datasets.
	
	From Figure \ref{fig:sharing}, we observe similar phenomenons as in the isolated model: the defending result decreases dramatically in the first 5 iterations, and after around 10 iterations the defending result is close to what it will eventually converge to.
	However, different from the isolated model, now we can no longer guarantee that the defending result of the mixed strategies are close to $\opt_f$, the lower bound for $\opt_m$.
	As discussed above, one possible reason can be that $\opt_f$ is much smaller than $\opt_m$.
	Unfortunately, unless there is way to give a tighter lower bound for $\opt_m$, there is no way to find out whether the mixed strategy \algName($30$) returns is close-to-optimal or not.
	We believe that this would be an interesting topic to study, and we leave it as the future work.

	\section{Conclusion and Future Works}
	In this work, we study mixed strategies for security games with general threshold and quantify its advantage against pure strategies. 
	We show that it is \textsf{NP}-hard to compute the optimal mixed strategy in general and provide strong upper and lower bounds for the optimal defending result of mixed strategies in some specific scenarios.
	We also propose the {\algName} algorithm for the computation of mixed strategies.
	By performing extensive experiments on 6 real-world datasets, we demonstrate the effectiveness and efficiency of the {\algName} algorithm by showing that even with very small support sizes, the mixed strategy returned is able to achieve defending results that are close-to-optimal. 
	Regarding future work, we believe that it would be most interesting to derive tighter lower bounds for the optimal defending results in the resource sharing model.
	In addition, studying mixed strategies against contagious attacks~\citep{aaai/BaiLYW21} and imperfect attackers~\citep{ress/ZhangWZ21} are also interesting directions.

\bibliographystyle{acm/named}
\bibliography{ref}

\end{document}